\documentclass[aps,preprint,showpacs,showkeys]{revtex4}
\usepackage{bm}

\input{tcilatex}
\begin{document}

\title{Bound state of solution of Dirac-Coulomb problem with spatially
dependent mass}
\author{Eser Ol\u{g}ar$^{1}$, Hayder M. Dhahir$^{1}$, H. Mutaf$^{1}$}
\affiliation{Gaziantep University , Engineering of Physics Department, Gaziantep/TURKEY}
\email{olgar@gantep.edu.tr, rg\_s3@yahoo.com, hmutaf@gantep.edu.tr}
\date{\today }

\begin{abstract}
The bound state solution of Coulomb Potentials in Dirac equation is
calculated for position dependent mass function $M(r)$ within the framework
asymptotic iteration method (AIM). The eigenfunctions are derived in terms
of hypergeometric function using function generator equation of AIM.
\end{abstract}

\maketitle

\section{Introduction}

The solution of the relativistic Dirac equation for quantum mechanical
systems in both case of spatially dependent mass and constant mass plays an
important role in many branches of physics \cite%
{1,2,3,4,5,6,7,8,9,10,11,12,13,14,15}. Since the importance of investigation
of the Dirac equation with position dependent mass, there has been increased
a great interest on it \cite{5,6,7,8,9,10,11,12,13,14,15}. Dirac equation
with position-dependent mass for solvable potentials has been addressed by a
number of different methods \cite{16,17,18,19,20,21,22,23}.Besides these
methods, recently Ciftci et al \cite{24,241,25} proposed an asymptotic
iteration method (AIM) which draws the attention of a many researchers for
relativistic equations \cite{26,261,27,28,281,282,283}.

This method has the advantages of obtaining the solution of eigenvalue
problems without solving directly the differential equation. Dirac equation
with first order differential equation form with two dimensional (2D)
Coulomb potential for constant mass \cite{29,30,31,32,321,322,33} and for
spatially dependent mass \cite{34} has been solved by using AIM. M. Hamzavi
and collaborators \cite{34} have been considered the Coulomb potential
including a Coulomb-like tensor potential under pseudospin symmetry limit.

The main purpose of this study is to solve the bound state sollution of
Coulomb potential for position dependent mass of Dirac equation without
Coulomb-like tensor potential by considering the relation between vector and
scalar potential as $S(r)=V(r)(b-1)$ where $b$ is arbitrary parameter \cite%
{28,281}. The second section represents the formalism of Dirac equation with
position dependent mass. The asymptotic iteration method is introduced in
Section 3. The calculation of eigenvalues and eigenfunctions of Coulomb
potential are outlined in the subsequent section. The last section devotes
to conclusion.

\section{Formalism of the Dirac Equation}

The Dirac equation for a central field in 3-dimensions is written for
spherically symmetric vector $V(r)$ and $S(r)$ spherically symmetric scalar
potential by using the parameters $\hbar =c=1$ as%
\begin{equation}
E_{nl}\text{ }\Psi =[\sum_{j=1}^{3}\alpha _{j}p_{j}+\beta (m+S(r))+V(r)]\Psi
\label{1}
\end{equation}%
where $m$ is the mass of the particle, $S(r)$ is a spherically symmetric
scalar potential, $V(r)$ is a spherically symmetric vector potential, $%
\alpha $ and $\beta $ are the usual Dirac matrices satisfying
anticommutation relations, and $E_{nl}$ is the corresponding eigenvalues.
After some algebraic calculations, one obtains the following first-order
linear coupled differential equations (\ref{1}) .

Without any approximation, the Dirac equation for a central field in
spherical coordinates can be separated into the variables. Thus, it has mean
the eigenfunction of the orbital and spin angular momentum can be found as

\begin{eqnarray}
\frac{dF_{nk}(r)}{dr}+\frac{k}{r}F_{nk}(r)
&=&(E_{nl}+M(r)-V(r)+S(r))Q_{nk}(r)  \label{2} \\
\frac{dQ_{nk}(r)}{dr}-\frac{k}{r}Q_{nk}(r)
&=&-(E_{nl}-M(r)-V(r)-S(r))F_{nk}(r)  \label{3}
\end{eqnarray}%
where $k=-(l+1)$ for the total angular momentum $j=l+1/2$, and $l$ is
angular momentum quantum number. $F_{nk}(r)$ and $Q_{nk}(r)$ are the radial
wave function of the upper and the lower-spinor components respectively, and
the general form of two second-order differential equations for
corresponding eigenfunctions are obtain by eliminating wave function $%
F_{nk}(r)$ in Eq. (\ref{2}) and $Q_{nk}(r)$ in Eq. (\ref{3}) we get%
\begin{equation}
\lbrack \frac{d^{2}}{dr^{2}}-\frac{k(k+1)}{r^{2}}]F_{nk}(r)-\frac{(\frac{%
dM(r)}{dr}-\frac{d\Delta (r)}{dr})(\frac{d}{dr}+\frac{k}{r})F_{nk}(r)}{%
M(r)+E_{nl}-\Delta (r)}=[(M(r)+E_{nl}-\Delta (r))(M(r)-E_{nl}+\tsum
(r))]F_{nk}(r)  \label{4}
\end{equation}%
and%
\begin{equation}
\left[ \frac{d^{2}}{dr^{2}}-\frac{k(k-1)}{r^{2}}\right] Q_{nk}(r)-\frac{(%
\frac{dM(r)}{dr}+\frac{d\Delta (r)}{dr})(\frac{d}{dr}-\frac{k}{r})Q_{nk}(r)}{%
M(r)-E_{nl}+\Delta (r)}=[(M(r)+E_{nl}-\Delta (r))(M(r)-E_{nl}+\tsum
(r))]Q_{nk}(r)  \label{5}
\end{equation}%
where

\[
\tsum (r)=V(r)+S(r),\text{ and }\Delta (r)=V(r)-S(r). 
\]

We use the relationship between scalar and vector potentials,$\ $%
\[
\ S(r)=V(r)(b-1),b\geq 0 
\]%
to define $\tsum (r)$ and $\Delta (r)$. In this general description of
scalar potential, by choosing $b$ parameter 0, 1, and 2 the scalar potential
leads to case of $S(r)=-V(r)$, $S(r)=0$ (purely vector potential), and $%
S(r)=V(r)$ respectively. The other choices of $b$ leads to the required
condition for the case of $S(r)>V(r)$. This transformation yields to

$\ $%
\[
\tsum (r)=bV(r),\text{ and }\Delta (r)=(2-b)V(r). 
\]

\section{Asymptotic Iteration Method}

The AIM is proposed to solve the second-order linear differential equation
in the form of%
\begin{equation}
y^{\prime \prime }=\lambda _{0}(x)y^{\prime }+s_{0}(x)y  \label{6}
\end{equation}%
where the functions $\lambda _{0}(x)$ and $s_{0}(x)$ are differentiable with 
$\lambda _{0}(x)\neq 0$. Shortly, the solution of any differential equation
can be written in form of Eq. (\ref{6}) has a general solution within the
framework AIM as

\[
y(x)=\exp (-\tint \alpha dt)[C_{2}+C_{1}\tint \exp (\tint (\lambda _{0}(\tau
)+2\alpha (\tau ))d\tau )dt] 
\]%
where $C_{i}$ are integral constants. The arbitrary functions for the limit
of $n$ are

\begin{eqnarray*}
\ \ \ \lambda _{n}(x) &=&\lambda _{n-1}^{\prime }(x)+s_{n-1}(x)+\lambda
_{0}(x)\lambda _{n-1}(x) \\
s_{n}(x) &=&s_{n-1}^{\prime }(x)+s_{0}(x)\lambda _{n-1}(x).
\end{eqnarray*}%
with asymptotic expression$\ $%
\[
\ \frac{\ s_{n}(x)}{\lambda _{n}(x)}=\frac{s_{n-1}(x)}{\lambda _{n-1}(x)}%
=\alpha (x) 
\]%
And the termination condition is in the form of

\begin{equation}
\Delta _{k}(x)=\left\vert 
\begin{array}{cc}
s_{n}(x) & \ \lambda _{n}(x) \\ 
s_{n-1}(x) & \ \lambda _{n-1}(x)%
\end{array}%
\right\vert =\lambda _{n-1}(x)s_{n}(x)-\lambda _{n}(x)s_{n-1}(x),\text{\quad 
}k=1,2,3,.\ldots  \label{8}
\end{equation}%
which gives the solution of physical systems.\bigskip

\section{Solution of the Dirac - Coulomb Problem}

The Dirac-Coulomb potential is considered by proposing the coulomb like
vector potential and scalar potential as

\begin{equation}
V(r)=\frac{V_{0}}{r}\text{, yields to }S(r)=\frac{V_{0}(b-1)}{r}\text{.}
\label{9}
\end{equation}%
The Eq. (\ref{4}) can not be solved analytically because of the effect of
the last term $(\frac{dM(r)}{dr}-\frac{d\Delta (r)}{dr})$. Therefore, we
calculate the required mass function that satisfies the equality $(\frac{%
dM(r)}{dr}-\frac{d\Delta (r)}{dr}=0)$ to eliminate this effect. Thus, using
this equality condition, the mass function is obtained as the following
function

\begin{equation}
M(r)=\frac{(2-b)V_{0}}{r}+m_{0}  \label{10}
\end{equation}%
where $m_{0}$ is the rest mass of the fermionic particle and $(2-b)V_{0}$ is
the perturbed mass \cite{35}. By substituting the potential functions in Eq.
(\ref{9}) and variable mass function in Eq. (\ref{10}) into Eq. (\ref{4}),
we get

\begin{equation}
\frac{-k(k+1)}{r^{2}}F_{nk}(r)-(E_{nl}+m_{0})(-E_{nl}+\frac{%
2V_{0}(E_{nl}+m_{0})}{r}+m_{0})F_{nk}(r)+\frac{d^{2}}{dr^{2}}F_{nk}(r)=0
\label{11}
\end{equation}%
At this point, if the following notations are made

\begin{equation}
-E_{nl}^{2}+m_{0}^{2}=\varepsilon _{nl}^{2},k(k+1)=A(A+1),\
-2(E_{nl}+m_{0})V_{0}=B,  \label{e11}
\end{equation}%
the eigenvalues equation transforms to

\begin{equation}
(-\varepsilon _{nl}^{2}-\frac{A(A+1)}{r^{2}}+\frac{B}{r})F_{nk}(r)+\frac{%
d^{2}F_{nk}(r)}{dr^{2}}=0  \label{12}
\end{equation}%
Propose the wavefunction by using AIM as

\begin{equation}
F_{nk}(r)=r^{A+1}\exp [-\varepsilon _{nl}]\chi (r)  \label{13}
\end{equation}%
Then, by substituting this wavefunction into Eq. (\ref{12}), we obtain

\[
(B-2(1+A)\varepsilon _{nl})\chi (r)+2(1+A-\varepsilon _{nl}r)\chi ^{\prime
}(r)+r\chi ^{\prime \prime }(r)=0. 
\]%
Thus, $\chi ^{\prime \prime }(r)$ becomes

\begin{equation}
\chi ^{\prime \prime }[r]=\frac{-2(1+A-\varepsilon _{nl}r)}{r}\chi ^{\prime
}[r]+\frac{-B+2(1+A)\varepsilon _{nl}}{r}\chi \lbrack r]  \label{14}
\end{equation}%
By comparing Eq. (\ref{14}) with the second-order differential equation of
Eq. (\ref{6}), we get the arbitrary functions $\lambda _{0}(r)$ and $%
s_{0}(r) $. The values of arbitrary functions are%
\begin{eqnarray}
\lambda _{0}(r) &=&2\left[ \frac{\varepsilon _{nl}r-A-1}{r}\right] , 
\nonumber \\
s_{0}(r) &=&\frac{2(A+1)\varepsilon _{nl}-B}{r}  \nonumber \\
\lambda _{1}(r) &=&\frac{4\varepsilon _{nl}^{2}r^{2}-6A\varepsilon
_{nl}r-6\varepsilon _{nl}r-Bu+4A^{2}+10A+6}{u^{2}},  \label{15} \\
s_{1}(r) &=&\frac{(B-2(A+1)\varepsilon _{nl})(2A-2\varepsilon _{nl}r+3)}{%
r^{2}}  \nonumber \\
&&..............................  \nonumber
\end{eqnarray}

By using termination condition for energy, we get%
\begin{eqnarray*}
\frac{s_{0}(r)}{\lambda _{0}(r)} &=&\frac{s_{1}(r)}{\lambda _{1}(r)}%
\Longrightarrow \varepsilon _{0l}=\frac{B}{2(A+1)} \\
\frac{s_{1}(r)}{\lambda _{1}(r)} &=&\frac{s_{2}(r)}{\lambda _{2}(r)}%
\Longrightarrow \varepsilon _{1l}=\frac{B}{2(A+2)} \\
\frac{s_{2}(r)}{\lambda _{2}(r)} &=&\frac{s_{3}(r)}{\lambda _{3}(r)}%
\Longrightarrow \varepsilon _{2l}=\frac{B}{2(A+3)} \\
&&...............
\end{eqnarray*}%
The general formula of $\xi $\ for n values can be written as%
\begin{equation}
\varepsilon _{nl}=\frac{B}{2(n+A+1)},\qquad n=0,1,2,.......  \label{e16}
\end{equation}%
The eigenvalues in Eq. (\ref{e16}) is transformed into the form of $E_{n}$
by the definition of the parameter $\varepsilon _{nl}$ in Eq. (\ref{e11})%
\begin{equation}
E_{nl}^{2}=m_{0}^{2}-\left( \frac{B}{2(n+A+1)}\right) ^{2}  \label{17}
\end{equation}

The corresponding energy eigenfunctions can be found by using the generator

\begin{equation}
\chi (r)=exp(-\int\limits^{r}\frac{s_{k}(r)}{\lambda _{k}(r)}dr  \label{19}
\end{equation}%
By applying the function generator, the $f_{n}(r)$ functions can be written
in series expansion by hypergeometric functions with constant $(B+n+1)^{n}$
and $\prod\limits_{k=0}^{(n-1)}(B+2+k)$. By generalizing these expansions,
we get%
\begin{equation}
\chi (r)=(B+n+1)^{n}\left[ \prod\limits_{0}^{n-1}(2B+2+k)\right]
x_{1}F_{1}(-n,2B+2;2\varepsilon _{nl}r)  \label{20}
\end{equation}%
We write the upper spinor component of the radial wave function as

\begin{equation}
F_{nk}(r)=r^{A+1}e^{(-\varepsilon _{nl}r)}(B+n+1)^{n}\left[
\prod\limits_{0}^{n-1}(2B+2+k)\right] x_{1}F_{1}(-n,2B+2;2\varepsilon _{nl}r)
\label{21}
\end{equation}%
The lower spinor wave function can be obtained in a similar algebraic
calculation. The mass function for lower spinor is calculated as $M(r)=\frac{%
(b-2)V_{o}}{r}+m_{0}$. After all same algebraic procedure, we get the same
results for eigenfunctions and eigenvalues with different parameters as%
\[
G_{nk}(r)=r^{A+1}e^{(-\varepsilon _{nl}r)}(B+n+1)^{n}\left[
\prod\limits_{0}^{n-1}(2B+2+k)\right] x_{1}F_{1}(-n,2B+2;2\varepsilon
_{nl}r) 
\]%
where%
\begin{eqnarray}
E_{nl}^{2}-m_{0}^{2} &=&-\varepsilon _{nl}^{2},  \label{23} \\
k(k-1)+4V_{0}^{2}(b^{2}-3b+2) &=&A(A+1),  \label{24} \\
2(E_{nl}+(2b-3)m_{0})V_{0} &=&B  \label{25}
\end{eqnarray}

These eigenfunctions and eigenvalues obey the results in \cite{34} after
mapping the corresponding parameters with those of in \cite{34}.

\section{Conclusion}

The spectrum of position dependent mass Dirac equation for Coulomb potential
is obtained within the framework of AIM method without solving the
differential equation. The mass function is considered as in the form of
function satisfying the equality condition, $\frac{dM(r)}{dr}-\frac{dV(r)}{dr%
}=0$. When the vector potential $V(r)$ is considered equal (states $b=2$) to
the spherical Scalar potential, the mass function is reduced to constant
mass situation. In upper spinor wavefunction, if $b>2$ $(S(r)>V(r))$, the
perturbed mass term results a negatively effect in $M(r)$. But in lower
spinor wavefunction, this condition yields to a positively effect in $M(r)$.
Therefore, by adjusting the parameter $b$, the bound-state solutions for
spinor wavefunctions is calculated by applying AIM and compared with
corresponding results in \cite{34}.

\section{Acknowledgement}

This research was supported by the Research Fund of Gaziantep University
(BAP) and the Scientific and Technological Research Council of TURKEY (T\"{U}%
B\.{I}TAK).

\end{document}